\documentclass[twocolumn,prl,showpacs,preprintnumbers,amssymb,nofootinbib]{revtex4-1}

\usepackage{epsfig}
\usepackage{dcolumn}
\usepackage{bm}

\usepackage{amsmath}
\usepackage{amssymb}
\usepackage{latexsym}
\usepackage{epsfig}
\usepackage{color}
\usepackage{mathtools}
\usepackage{youngtab}
\usepackage[normalem]{ulem}
\usepackage{siunitx}

\def\ln{{\lambda_N}}
\def\aln{{A_{\lambda_N}}}

\def\neutl{{{\tilde\chi}^0_{1}}}

\def\snr{{\tilde N}}

\def\snmassr{{m_{\tilde N_1}}}

\def\rhn{{N}}

\def\higgsl{{H_1^0}}
\def\higgsm{{H_2^0}}

\def\hmassl{m_{H_1^0}}
\def\hmassm{m_{H_2^0}}

\def\phiggsl{{A_1^0}}

\def\phmassi{{m_{A_a^0}}}
\def\phmassl{{m_{A_1^0}}}

\newcommand{\sigsi}{\sigma^{SI}_{\snr_1 p}}

\def\sigmav{\langle\sigma v\rangle_0}
\def\sigmavgg{\langle\sigma v\rangle_{0,\gamma\gamma}}

\newcommand{\crosssec}{\sigma_{\snr_1 p}^{\rm SI}}

\newcommand{\snrelic}{{\Omega_{\snr_1} h^2}}

\newcommand{\bsg}{b\to s\gamma}
\newcommand{\bmumu}{B_S\to\mu^+\mu^-}
\newcommand{\btaunu}{B^+ \to \tau^+ \nu_\tau}

\def\lsim{\raise0.3ex\hbox{$\;<$\kern-0.75em\raise-1.1ex\hbox{$\sim\;$}}}
\def\gsim{\raise0.3ex\hbox{$\;>$\kern-0.75em\raise-1.1ex\hbox{$\sim\;$}}}

\allowdisplaybreaks

\begin{document}

\preprint{IPPP/15/01}
\preprint{DCTP/15/02}
\preprint{IFT-UAM/CSIC-15-001}
\preprint{FTUAM-15-1}

\title{
Fits to the Fermi-LAT GeV excess with RH sneutrino dark matter: implications for direct and indirect dark matter searches and the LHC
}

\author{  D.G.~Cerde\~no$^a$, M.~Peir\'o$^{b,c}$, and S.~Robles$^{b,c}$}

\affiliation{\small\slshape  $^a$ Institute for Particle Physics Phenomenology, Department of Physics
Durham University, Durham DH1 3LE, United Kingdom\\
      $^b$ Instituto de F\'{\i}sica Te\'{o}rica UAM/CSIC, Universidad Aut\'{o}noma de Madrid,  28049, Madrid, Spain\\
      $^c$ Departamento de F\'{\i}sica Te\'{o}rica,
      Universidad Aut\'{o}noma de Madrid, 28049
      Madrid, Spain\\ }
\begin{abstract}

We show that the right-handed (RH) sneutrino in the NMSSM can account for the observed excess in the Fermi-LAT spectrum of gamma-rays from the Galactic Centre, while fulfilling all the current experimental constraints from the LHC as well as from direct and indirect dark matter searches.
We have explored the parameter space of this scenario, computed the gamma-ray spectrum for each phenomenologically viable solution and then performed a $\chi^2$ fit to the excess. 
Unlike previous studies based on model-independent interpretations, we have taken into account the full annihilation spectrum,  
without assuming pure annihilation channels.
Furthermore, we have incorporated limits from direct detection experiments, LHC bounds and also the constraints from Fermi-LAT on dwarf spheroidal galaxies (dSphs) and gamma-ray spectral lines. 
In addition, we have estimated the effect of the most recent Fermi-LAT reprocessed data (Pass~8). 
In general, we obtain good fits to the GCE when the RH sneutrino annihilates mainly into pairs of light singlet-like scalar or pseudoscalar Higgs bosons  that subsequently decay in flight, producing four-body final states and spectral features that improve the goodness of the fit at large energies. 
The best fit ($\chi^2=20.8$) corresponds to a RH sneutrino with a mass of 64~GeV which annihilates preferentially into a pair of light singlet-like pseudoscalar Higgs bosons (with masses of order 60 GeV).
Besides, we have analysed other channels that also provide good fits to the excess. 
Finally, we discuss the implications for direct and indirect detection searches paying special attention to the possible appearance of gamma-ray spectral features in near future Fermi-LAT analyses, as well as deviations from the SM-like Higgs properties at the LHC. 
Remarkably, many of the scenarios that fit the GCE can also be probed by these other complementary techniques. 
\end{abstract}
\maketitle

\section{Introduction}

A large number of cosmological and astrophysical observations have evidenced that 85\% of the matter content of the Universe is in the form 
of dark matter (DM). A generic weakly interacting massive particle (WIMP) is a well-motivated candidate for this new kind of matter, 
since its thermal production in the Early Universe would match the observed DM abundance. In addition, WIMPs can be easily accommodated in theories beyond the Standard Model, such as Supersymmetry (SUSY).

WIMPs can be searched for indirectly, through the particles produced when they annihilate in the DM halo (photons, neutrinos and antiparticles). 
Among the different annihilation products, gamma-rays provide an appealing detection possibility because the signal can be traced back to the source. 
The Large Area Telescope (LAT) aboard the Fermi Gamma-ray Space Telescope has produced the most detailed maps of the gamma-ray sky for a wide range of energies, with unprecedented angular and energy resolutions. 
Using data from the Fermi-LAT, various studies have revealed the presence of an excess from an extended gamma-ray source in the inner region of the Galaxy~\cite{Vitale:2009hr,Hooper:2010mq,Morselli:2010ty,Hooper:2011,Abazajian:2012pn,Daylan:2014rsa}, 
a signal that is robust when known uncertainties are taken into account~\cite{Gordon:2013vta,Abazajian:2014fta,Zhou:2014lva,Calore:2014xka}. 
Although the explanation of this Galactic Centre excess (GCE) is still under debate\footnote{A recent study shows that the GCE might be produced, within current uncertainties, by a population of unresolved millisecond pulsars~\cite{Petrovic:2014xra}.}, 
if it were interpreted in terms of DM annihilations \cite{Hooper:2010mq,Hooper:2011,Abazajian:2012pn,Daylan:2014rsa,Gordon:2013vta,Abazajian:2014fta,Zhou:2014lva,Calore:2014xka,Agrawal:2014oha,Calore:2014nla} it would correspond to a particle with a mass in the range $30-50$~GeV for a $b\bar{b}$ final state ($7-10$~GeV for a $\tau\bar{\tau}$ final state) and with an annihilation cross section in the DM halo, $\sigmav \sim 1-2\times 10^{-26}$~cm$^3$/s, remarkably close to that expected from a thermal relic.

Several attempts have been made to explain the GCE in terms of simplified models for DM \cite{Belikov:2010yi,Buckley:2010ve,Zhu:2011dz,Marshall:2011mm,Buckley:2011vs,Boucenna:2011hy,Buckley:2011mm,Hooper:2012cw,Cotta:2013jna,Buckley:2013sca,Hagiwara:2013qya,Okada:2013bna,Fortes:2013ysa,Alves:2013tqa,Modak:2013jya,Boehm:2014hva,Alvares:2012qv,Boehm:2014hva,Alves:2014yha,Boehm:2014bia,Izaguirre:2014vva,Abdullah:2014lla,Berlin:2014tja,Ghosh:2014pwa,Martin:2014sxa,Basak:2014sza,Ko:2014gha,Berlin:2014pya,Wang:2014elb,Arina:2014yna,Okada:2014usa,Bell:2014xta,Banik:2014eda,Calore:2014nla,Cheung:2014tha,Hooper:2014fda,Liu:2014cma,Dolan:2014ska,Biswas:2014hoa,Ghorbani:2014gka}, 
considering DM annihilation into pure channels. However, as pointed out in Ref.~\cite{Cahill-Rowley:2014ora}, it is crucial to investigate if this excess can be obtained within a complete theoretical framework. 
For example, it has been recently shown that the neutralino could reproduce the GCE for DM masses up to hundreds of GeV depending on the primary annihilation channel within the context of the MSSM \cite{Agrawal:2014oha} and the Next-to-MSSM (NMSSM) \cite{Cheung:2014lqa,Huang:2014cla,Gherghetta:2015ysa}.

In this article, we carry out a complete analysis of the right-handed (RH) sneutrino in the NMSSM \cite{Cerdeno:2008ep,Cerdeno:2009dv} and demonstrate that
it can successfully account for the GCE while fulfilling all the experimental constraints from direct and indirect DM searches as well as collider physics. 
We apply the LUX and SuperCDMS limits on the spin-independent elastic scattering cross section of DM off protons, which are currently the most stringent bounds from direct detection experiments. 
We also consider the latest results from the LHC on the Higgs boson mass and couplings to the SM particles, which are known to be specially constraining for light DM scenarios through the upper bound on the invisible and non Standard Model Higgs decays. 
Besides, the latest bounds from the measurement of the rare decays $\bmumu$, $\bsg$ and $\btaunu$ are also applied. 
Finally, we incorporate the Fermi-LAT constraints on dwarf spheroidal galaxies (dSphs) and spectral feature searches in the gamma-ray spectrum, 
including an estimation of the effect that the most recent results derived from the Pass 8 data impose on our results.

\begin{figure*}
      \begin{center}
	\scalebox{0.9}{
            \raisebox{5ex}{\includegraphics[scale=0.4]{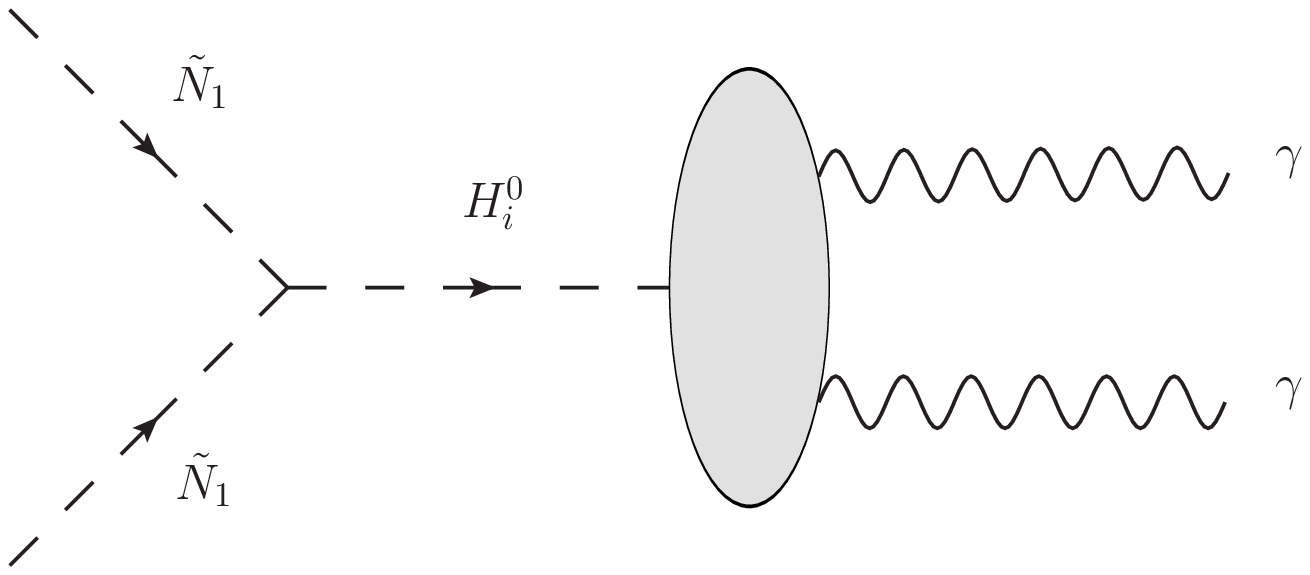}}
            \includegraphics[scale=0.4]{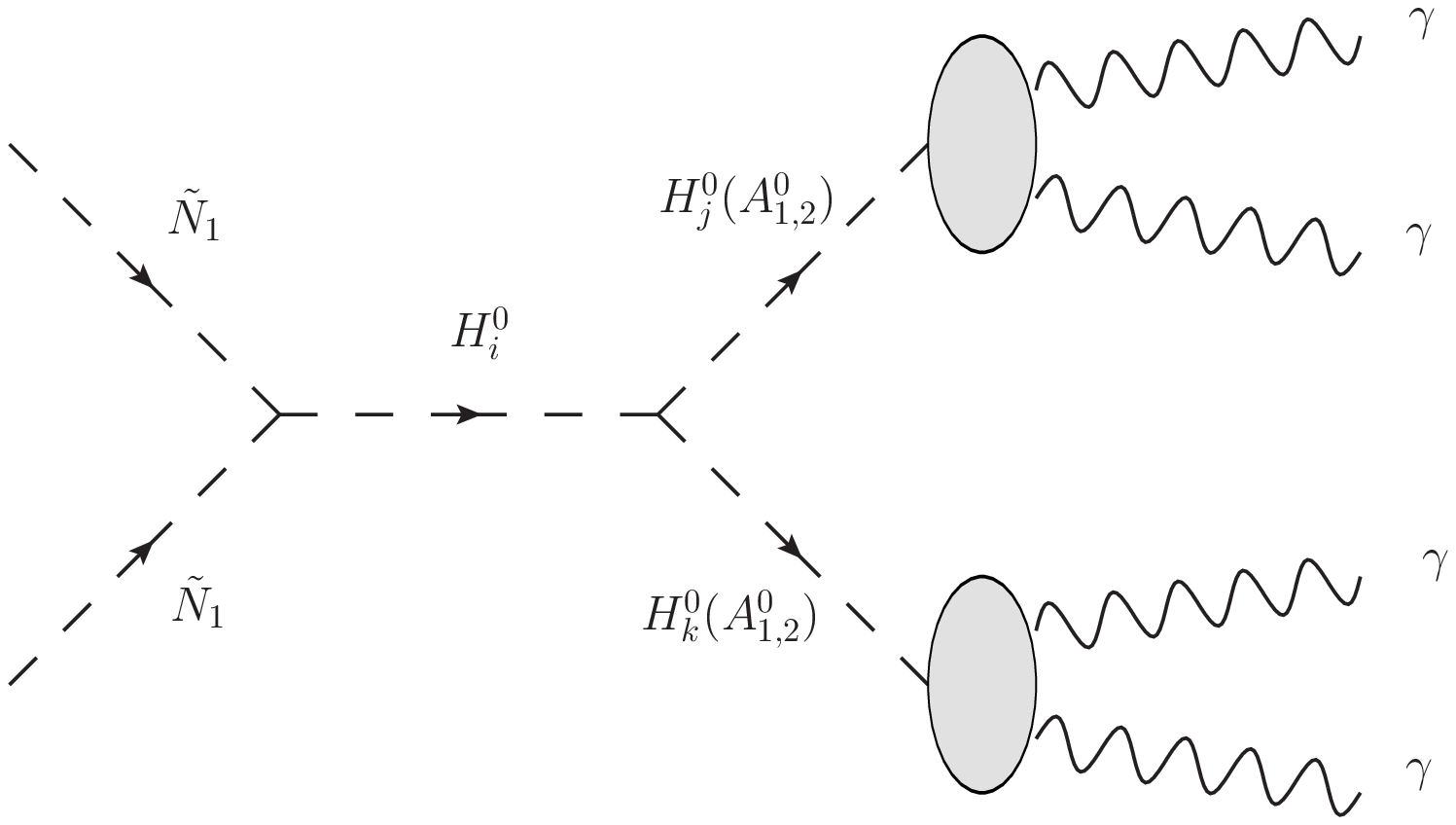}
            \includegraphics[scale=0.4]{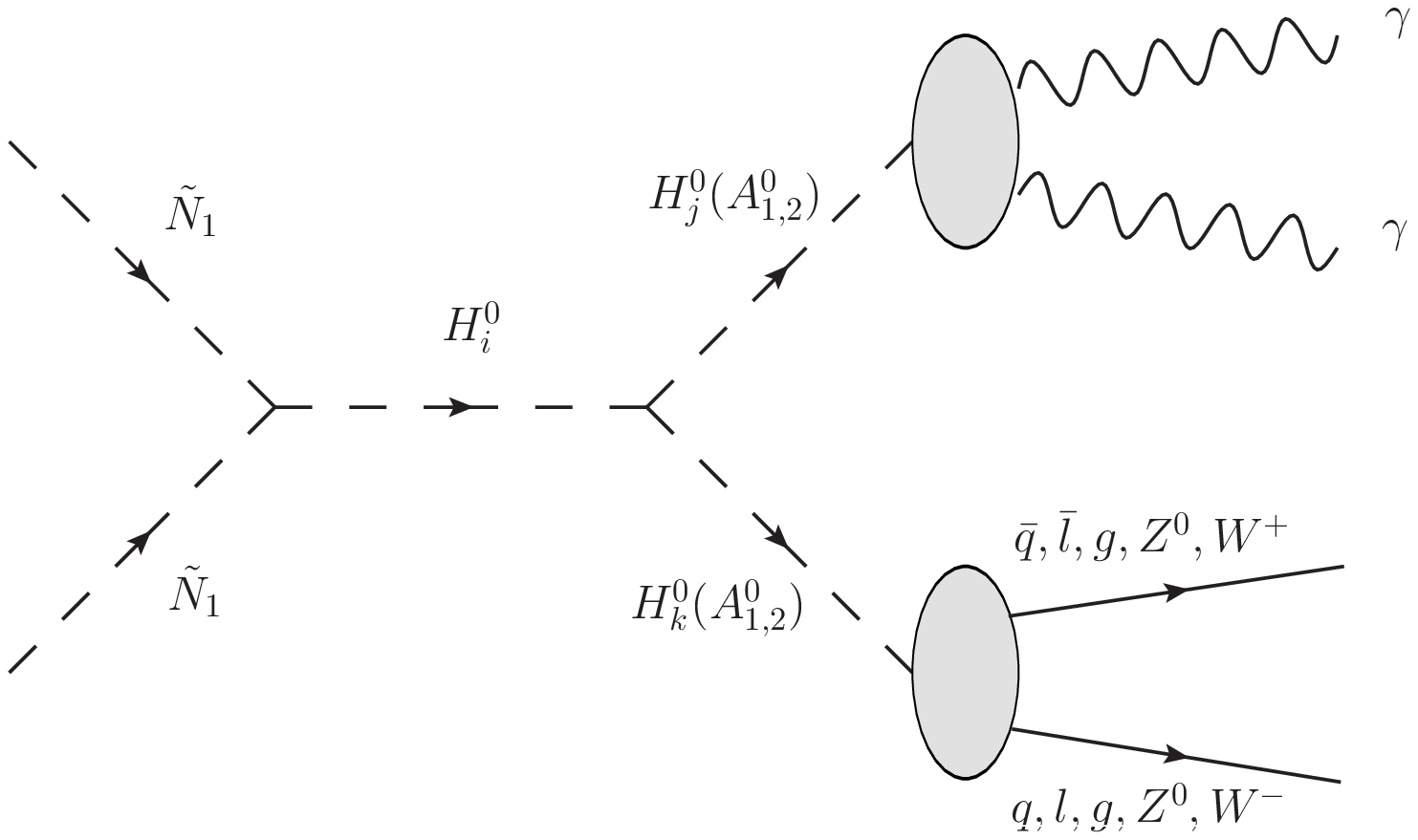}
               }
	\caption{\footnotesize RH sneutrino annihilation diagrams that produce lines and box-shaped features in the gamma-ray spectrum.}
      	\label{fig:diagram}
      \end{center}
\end{figure*}

\section{Fitting the GCE with RH sneutrinos}

This model has been extensively described in Refs.~\cite{Cerdeno:2008ep,Cerdeno:2009dv}. 
It is an extended version of the NMSSM, in which a new gauge singlet superfield, $N$, is introduced in order to account for RH neutrino and sneutrino states as in \cite{Kitano:1999qb,Garbrecht:2006az}. The superpotential of this construction is given by
\begin{eqnarray}
  W &=& W_{\rm NMSSM} + \lambda_N S N N + y_N L \cdot H_2 N, 
    \label{eq:superpotential}
\end{eqnarray}
where flavour indices are omitted and the dot denotes the antisymmetric $SU(2)_L$ product. $ W_{\rm NMSSM}$ is the NMSSM superpotential, $\ln$ is a new dimensionless coupling, 
$y_N$ is the neutrino Yukawa coupling, and $H_{1,2}$ are the down and up type doublet Higgs components, respectively.
As in the NMSSM, a global $Z_3$ symmetry is imposed so that there are no supersymmetric mass terms in the superpotential.
Since we assume $R$-parity conservation in order to guarantee the stability of the LSP, the terms $NNN$ and $SSN$ are forbidden. Furthermore, we do not consider CP violation in the Higgs sector.

After radiative electroweak symmetry breaking the Higgs fields get non-vanishing vacuum expectation values (VEVs) and the physical Higgs states correspond to a superposition of the $H_d$, $H_u$ and $S$ fields. 
The RH sneutrino interacts with the SM particles through the mixing in the Higgs sector thanks to the coupling $\lambda_N S N N$, thereby behaving as a WIMP.

Interestingly, light RH sneutrinos with masses in the range of $10-150$~GeV are viable DM particles \cite{Cerdeno:2011qv} and constitute ideal candidates to account for the GCE, as we already pointed out in Ref.\,\cite{Cerdeno:2014cda}. 
Their phenomenology is very rich, as they can annihilate into a variety of final states, some of which include scalar and pseudoscalar Higgses. 
In particular, if $\snmassr>m_{H^0_1(A^0_1)}$, the annihilation final state of RH sneutrinos is dominated by a $H^0_1H^0_1(A^0_1A^0_1)$ pair in vast regions of the parameter space. 
The subsequent decay of each scalar and pseudoscalar Higgs into pairs of fermions or gauge bosons gives rise to non-standard final states, which often display spectral features coming from the $\gamma\gamma$ final states.

Given that the RH sneutrino annihilation contains a mixture of final states, often including exotic configurations, the GCE model-independent approach generally found in the literature is not applicable. 
To fit the GCE we have followed the approach of Ref.~\cite{Calore:2014xka}, where the authors take into account theoretical model systematics by exploring a large range of Galactic diffuse emission models\footnote{When these model systematics are included as correlated errors in the residual spectrum, the best fit for the DM interpretation of the model-independent analysis of Ref.~\cite{Calore:2014xka} is obtained for a $b\bar{b}$ final state with a mass of 49$^{+6.4}_{-5.4}$ GeV and a velocity averaged annihilation cross section of $1.76^{+0.28}_{-0.27}\times10^{-26}$ cm$^3$/s. 
Other analyses of the GCE employ different assumptions on the Galactic diffuse and point source components, and the reconstructed DM mass and annihilation cross section differ slightly.}. 
Regarding the DM distribution, we have considered a generalised NFW profile with slope $\gamma=1.2$ in the region of interest (ROI) $2^{\circ}\leq|b|\leq20^{\circ}$ and $|l|\leq20^{\circ}$, as in Ref.~\cite{Calore:2014xka}.

To implement the aforementioned analysis, we have performed a series of scans over the parameter space of the model 
in the RH sneutrino mass range $1-150$~GeV,
computing the gamma-ray spectrum as well as the RH sneutrino relic abundance with {\tt micrOMEGAs 3.6.9}~\cite{Belanger:2013oya}. 
The number of free parameters and the corresponding ranges of variation coincide with those used in Ref.~\cite{Cerdeno:2014cda} (only the range in the RH sneutrino mass has been enlarged to accommodate new solutions). In our analysis, we have also incorporated the most recent constraints from direct detection experiments using the latest results of LUX and SuperCDMS \cite{Akerib:2013tjd,Agnese:2013jaa,Agnese:2014aze}. 
LHC constraints on the masses of supersymmetric particles, the mass and couplings of the SM Higgs boson, as well as bounds from the rare decays $\bmumu$, $\bsg$ and $\btaunu$, have also been implemented (for more details, see Ref.~\cite{Cerdeno:2014cda}).

We have set an upper bound on the RH sneutrino relic abundance, $\Omega_{\snr_1} h^2<0.13$,
consistent with the latest Planck results~\cite{Ade:2013zuv}. 
Besides, we have considered the possibility that RH sneutrinos only contribute to a fraction of the total relic density and set for concreteness a lower bound on the relic abundance, $0.001<\Omega_{\snr_1} h^2$. 
To deal with these cases, the fractional density, $\xi=\min[1,\Omega_{\snr_1}h^2/0.11]$, has been introduced to account for the reduction in the rates for direct and indirect searches (assuming that the RH sneutrino is present in the DM halo in the same proportion as in the Universe).

Then, we have convoluted the differential photon spectrum of the scan points fulfilling all these constraints with the energy resolution of the LAT instrument. We have used the {\tt P7REP-SOURCE-V15} total (front and back) resolution of the reconstructed incoming photon energy as a function of the energy for normally incident photons.

Afterwards, we have calculated the $\chi^2$ function as follows~\cite{Calore:2014xka}:
\begin{equation}
	\chi^2= \sum_{ij}\left(\frac{d\bar{N}}{dE_i}(\boldsymbol{\theta})-\frac{dN}{dE_i}\right)
	\Sigma^{-1}_{ij}\left(\frac{d\bar{N}}{dE_j}(\boldsymbol{\theta})-\frac{dN}{dE_j}\right),
	\label{eq:chi2}
\end{equation}
where $\Sigma_{ij}$ is the covariance matrix containing the statistical errors and the diffuse model and residual systematics~\cite{Calore:2014xka}. $dN/dE_i$ ($d\bar{N}/dE_i$) stands for the measured (predicted) flux in the $i$th energy bin. 
The vector $\boldsymbol{\theta}$ refers to all parameters of our model which determine the predicted flux.

\subsection{Constraints from indirect DM searches}

The Fermi-LAT satellite has also provided bounds on the DM annihilation cross section in the Galactic halo derived from the study of the gamma-ray spectrum 
from dSphs and the search for spectral features in the Galactic Centre. 
These limits play an important role in the current analysis. Let us review them in more detail. 
\\[1.ex]

\noindent $\bullet$ Dwarf spheroidal galaxies

\noindent
The mass of these objects is dominated by DM; hence, they constitute ideal targets for indirect searches. 
The Fermi-LAT Collaboration has performed an analysis of the gamma-ray emission from 25 dSphs using four years of data \cite{Ackermann:2013yva}. 
The absence of a signal can be interpreted as constraints on the annihilation cross section of DM particles. 
It is customary to assume annihilation into pure SM channels in the calculation of these bounds.

The occurrence of non standard annihilation final states in our model prevents us from using these results directly. Instead, 
we have extracted independent upper bounds on $\xi^2\sigmav$ for each of the six more constraining dSphs (Coma Berenices, Draco, Segue I, Ursa Major II, Ursa Minor and Willman I), using the DM flux predicted by our model and the mean values for the $J$ factors from Ref.~\cite{Ackermann:2013yva}. 
Then, we have applied the most restrictive of these limits to our data.  
We have checked that this method leads to slightly less stringent bounds than the combined limit from the Fermi-LAT Collaboration (by a factor smaller than $1.5$), 
when applied to the region of DM masses from 10 to 100 GeV with pure annihilation channels.

Lastly, we have also estimated the impact of the preliminary results derived from the latest data (Pass~8) presented by the Fermi-LAT Collaboration~\cite{Ackermann:2015zua}. 
In general for any final state, the limit on $\xi^2\sigmav$ improves by approximately a factor $4-5$ for a DM mass in the range $10-130$~GeV. 
Conservatively, we have used a factor 4 to assess the dSph bounds derived from the newest Fermi data.
\\[1.ex]

\begin{figure*}
      \begin{center}
\scalebox{0.9}{
            \includegraphics[scale=1]{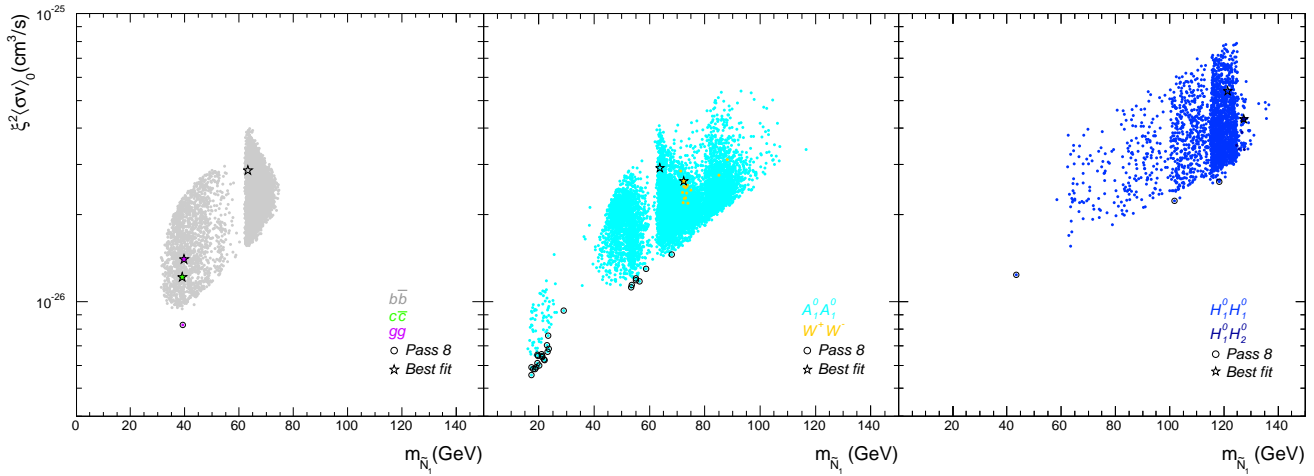}
               }
\caption{\footnotesize Velocity-averaged annihilation cross section of RH sneutrinos in the Galactic halo as a function of the RH sneutrino mass. All the points provide a fit to the GCE at 95\% C.L. 
We have also imposed all the experimental bounds, including dSph constraints and Fermi-LAT searches for spectral lines. The colour code indicates different dominant final states of the RH sneutrino annihilation. The best fit points for each annihilation channel are represented by a star. Points allowed by our estimation of the Pass~8 dSph bounds are encircled in black. }
      \label{fig:sv95P7}
      \end{center}
\end{figure*}

\noindent $\bullet$ Gamma-ray spectral features

The Fermi-LAT Collaboration has performed a search for spectral lines in the energy range $5-300$~GeV \cite{Ackermann:2013uma}. 
Not having found  any globally significant spectral feature, this analysis has been translated into 95\% C.L. upper limits on the DM annihilation cross section into a pair of photons, $\sigmavgg$.

The RH sneutrino in the NMSSM can give rise to a complex spectrum, displaying lines and box-shaped spectral features\footnote{ For a more detailed discussion on these features see Refs.~\cite{Ibarra:2012dw,Ibarra:2013eda}.}, which arise from the diagrams shown in Figure~\ref{fig:diagram}, and involve annihilation into pairs of light scalar and pseudoscalar Higgs bosons. 
The first diagram shows the usual contribution to the primary production of a pair of photons through a loop of charginos, sfermions, top quarks, $W^{\pm}$ and charged Higgses, and would produce a line with $E_\gamma=m_{\tilde{N}_1}$. 
The second and third diagrams would produce a line with energy $E_\gamma=m_{H^0_{j,k}}/2$ or $E_\gamma=m_{A^0_{1,2}}/2$ if any of the Higgs bosons were produced nearly at rest, when 
$2\,m_{\tilde{N}_1}\approx m_{H^0_j}+m_{H^0_k}$ or $2\,m_{\tilde{N}_1}\approx m_{A^0_j} + m_{A^0_k}$.
Otherwise, the decay in flight of the boosted Higgs bosons gives rise to box-shaped features with a maximum energy $E_\gamma^{max}=m_{\tilde{N}_1}/2$ and 
widths $\Delta E_{\gamma}=m_{\tilde{N}_1} - m_{H^0_{j,k}}(m_{A^0_{1,2}})$.

The published bounds \cite{Ackermann:2013uma} on $\sigmavgg$ do not include the specific DM halo used in this paper for the analysis of the GCE. In order to recalculate this limit, we have computed the $J$ factor for our halo in the region of interest (ROI) 
R41\footnote{This ROI is defined as a 41$^{\circ}$ circular region centred on the Galactic centre with a mask applied to $|b|<5^{\circ}$ and $|l|>6^{\circ}$, and has been optimised for a regular ($\gamma=1$) NFW profile.}, 
$J_{GCE}=13.0\times 10^{22}$~GeV$^2$~cm$^{-5}$, and compared it with the one used in Ref.\,\cite{Ackermann:2013uma}, which yields 
$J_{LAT}=8.53\times 10^{22}$~GeV$^2$~cm$^{-5}$. 
The ratio $r=J_{LAT}/J_{GCE}\approx1.52$ is then applied to the Fermi-LAT bounds on $\sigmavgg$ from Ref.~\cite{Ackermann:2013uma}.

Then, we have applied the bounds on the annihilation cross section into two photons to the $\xi^2\sigmavgg$ predicted by our model for monochromatic gamma-ray lines. 
Concerning the box-shaped contributions, first we have derived the corresponding limits on the annihilation cross section from the Fermi-LAT gamma-ray line 
bounds\footnote{The use of this derived bound for box-shaped features is well-motivated since the energy binning of the Fermi-LAT flux is chosen to be of the order of the energy resolution of the instrument. 
Hence, we are allowed to approximate this contribution as a continuum of lines extending from the minimum to the maximum box energy.} and 
afterwards we have applied them to our prediction weighted by the fractional DM density squared, $\xi^2$, along the box width. 
Finally, note that stronger bounds could be obtained if we used a ROI which is optimised for the profile used here, but this is out the scope of this article.

\section{Results}

\begin{figure*}
      \begin{center}
\scalebox{0.9}{
            \includegraphics[scale=0.5]{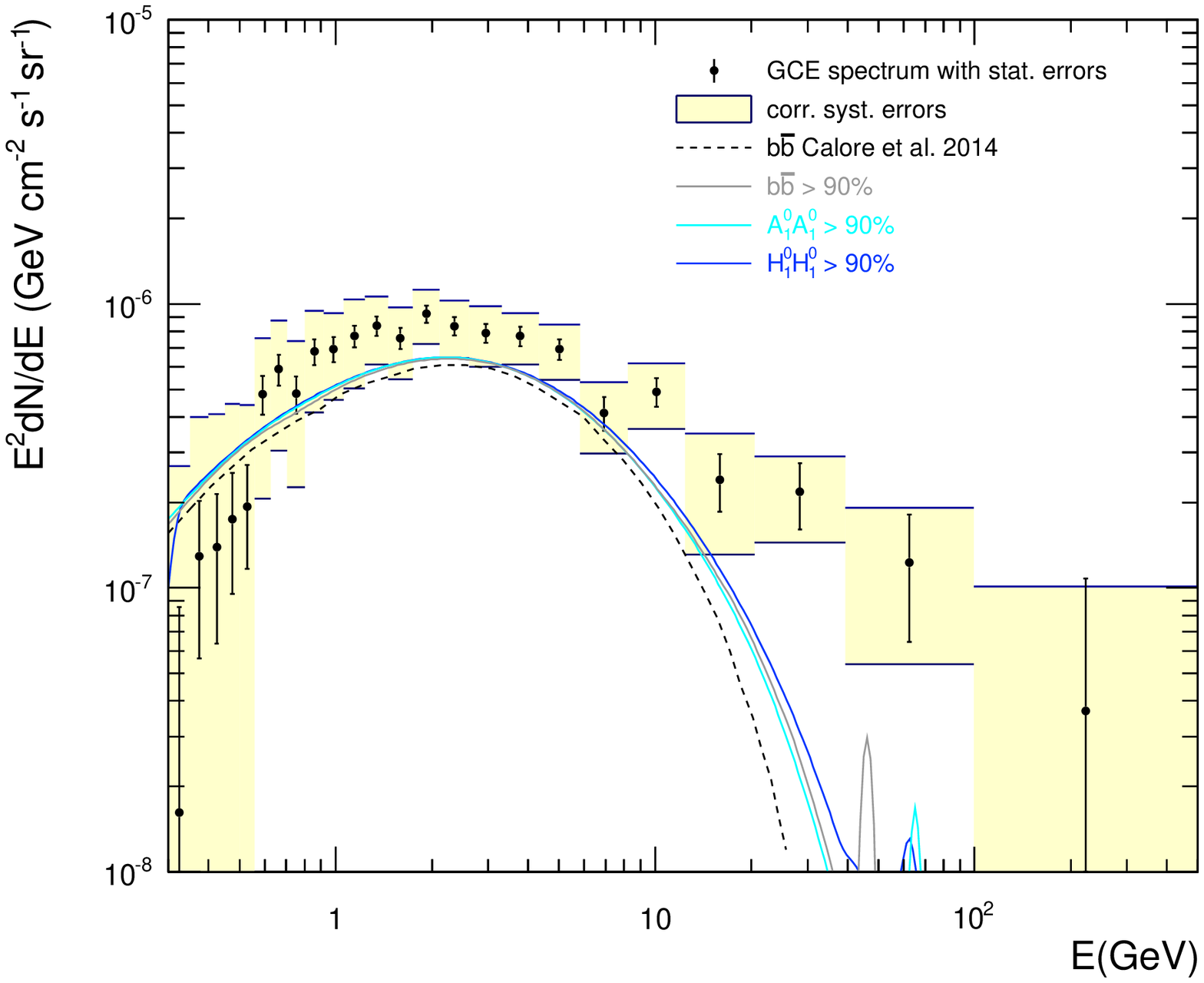}
            \includegraphics[scale=0.5]{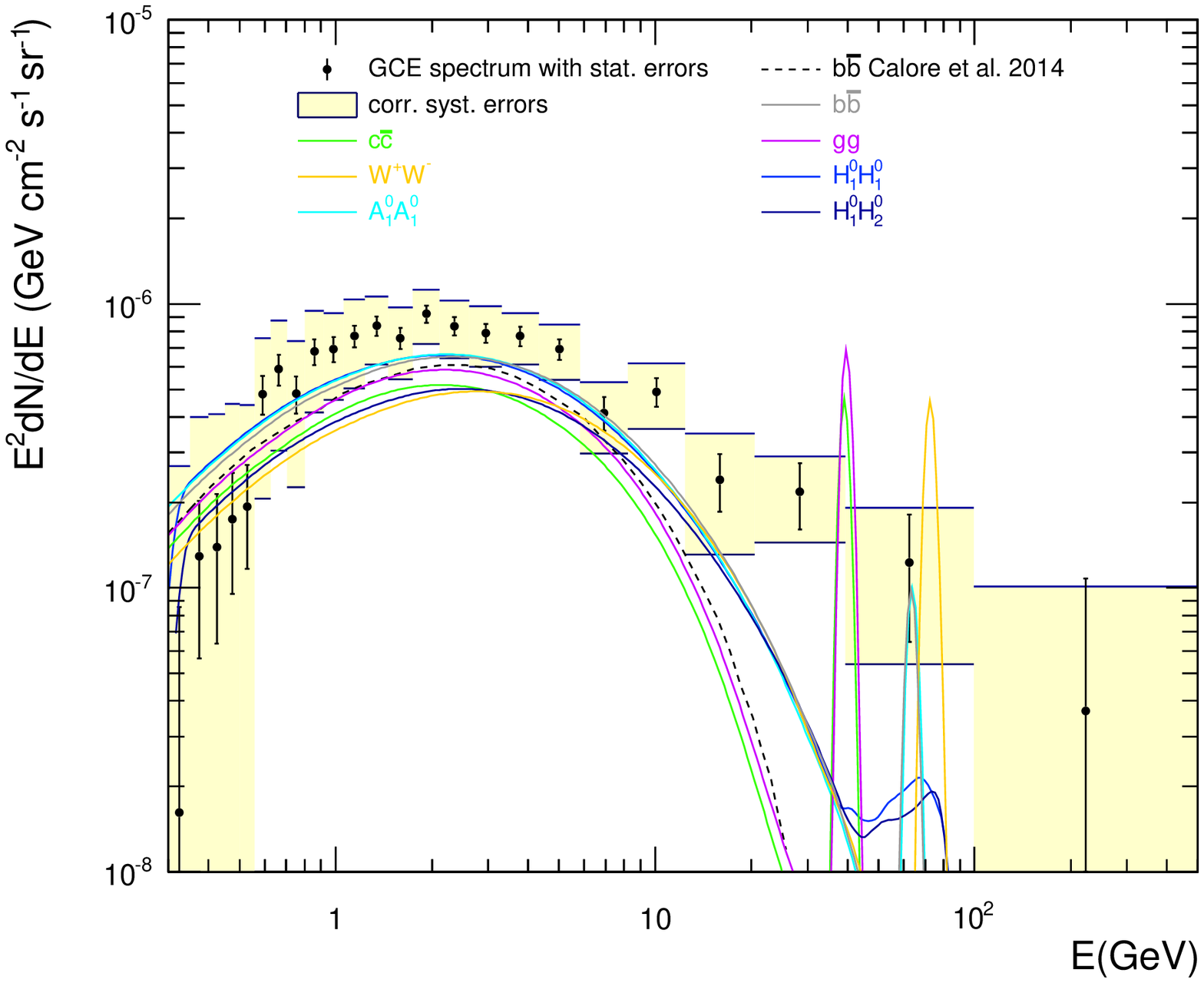}
               }
	\caption{\footnotesize Differential gamma-ray spectrum for the points in Table~\ref{tab:chi} for pure (left panel) and predominant (right 	panel) annihilation channels. The colour convention is as in Fig.\,\ref{fig:sv95P7}. The experimental data and errors are extracted from Ref.~\cite{Calore:2014xka}, as well as the best fit for a pure $b\bar{b}$ channel, represented by a black dashed line. }
      \label{fig:spec}
      \end{center}
\end{figure*}

In Figure \ref{fig:sv95P7}, we show $\xi^2\sigmav$ vs.  $\snmassr$ for the points of the parameter space that fit the GCE at 95\% C.L.
The different colours indicate the main annihilation channel (remember that the whole annihilation spectrum is considered when calculating the gamma-ray flux). 
The stars represent the best fit point for each of the dominant annihilation channels and their properties are summarised in Table\,\ref{tab:chi}, 
where we distinguish ``pure final states'' (if the main annihilation channel contributes to more than 90\% to $\xi^2\sigmav$) and ``mixed final states''.
Lastly, black circles correspond to the points that would be allowed by our estimation of the Pass~8 constraints on dSphs.

\begin{table}[t]
  \centering
  \begin{tabular}{|@{}c|S[table-format=3.1]|@{}c|@{}c|@{}c|}
    \multicolumn{5}{c}{Pure final states} \\[3pt]
    \hline
    Final state & {$\snmassr$ (GeV)} & $\xi^2\langle\sigma v\rangle_0$ (cm$^3$/s) & $\snrelic$ & $\chi^2$ \\[3pt]
    \hline
    $H^0_1H^0_1$ \hfill ($91.8$\%) & 119.8 & $5.1\times 10^{-26}$  & 0.094 & 21.9  \\[3pt]
    \hline
     $A^0_1A^0_1$ \hfill ($90.6$\%) & 65.0 & $2.7\times 10^{-26}$ & 0.109 & 22.3  \\[3pt]
    \hline
    $b\bar{b}$ \hfill ($90.2$\%) & 46.1 & $1.9\times 10^{-26}$ & 0.038 & 22.6  \\[3pt]
    \hline
    \multicolumn{5}{c}{Mixed final states} \\[3pt]
    \hline
    Final state & {$\snmassr$ (GeV)} & $\xi^2\langle\sigma v\rangle_0$ (cm$^3$/s) & $\snrelic $&$\chi^2$ \\[3pt]
    \hline 
     $A^0_1A^0_1$ \hfill ($44.7$\%) & 63.8 & $2.9\times 10^{-26}$ & 0.061 & 20.8  \\[3pt]
    \hline
     $b\bar{b}$ \hfill ($42.1$\%) & 63.2 & $2.9\times 10^{-26}$ & 0.042 & 21.0  \\[3pt]
    \hline
     $H^0_1H^0_1$ \hfill ($71.4$\%) & 121.4 & $5.4\times 10^{-26}$ & 0.075 & 21.6  \\[3pt]
    \hline
     $gg$ \hfill ($38.8$\%) & 39.6 & $1.4\times 10^{-26}$ & 0.071 & 23.7  \\[3pt]
    \hline
     $c\bar{c}$ \hfill ($33.0$\%) & 39.0 & $1.2\times 10^{-26}$ & 0.099 & 25.4  \\[3pt]
    \hline
     $H^0_1H^0_2$ \hfill ($44.5$\%) & 127.4 & $4.3\times 10^{-26}$ & 0.054 & 25.9  \\[3pt]
    \hline
     $A^0_1A^0_1$ $(4\tau)$ \hfill ($67.5$\%) & 25.5 & $1.5\times 10^{-26}$ & 0.068 & 27.4  \\[3pt]
    \hline    
    $W^+W^-$ \hfill ($28.0$\%) & 72.4 & $2.6\times 10^{-26}$ & 0.104 & 29.2  \\[3pt]
    \hline
  \end{tabular}
  \caption{Properties of the points that provide the best fit to the GCE for different annihilation final states. We have separated the solutions into \textit{pure} final states (which have an annihilation percentage into a given channel bigger than 90\%) and \textit{mixed} final states (in which case we show the dominant channel with its percentage). }
  \label{tab:chi}
\end{table}

As we can observe, there are solutions that fit the GCE for RH sneutrino masses in the range $\snmassr=15-135$~GeV, while fulfilling all other experimental constraints (from direct and indirect dark matter searches as well as from the LHC). 
The best fit points for pure annihilation channels are in good agreement with model-independent studies \cite{Agrawal:2014oha,Calore:2014xka}, but we have also obtained new non-standard annihilation channels (into light scalar and pseudoscalar singlet-like Higgs bosons)
and examples with mixed final states which provide a slightly better fit to the GCE.
Notice also that the points are separated in two regions in the RH sneutrino mass. Let us comment in more detail these two regions. 
\\[1ex]

\noindent $\bullet$ $\snmassr\approx15-30$~GeV. 
We have found solutions where the RH sneutrino annihilates mainly into a pair of very light, singlet-like, CP-odd Higgs bosons (cyan). Since $\phmassl<2m_b$, these pseudoscalars cannot decay into a pair of $b$ quarks and instead they do it predominantly into a pair of $\tau$ leptons. 
The resulting process, $\snr_1\snr_1\rightarrow 2\phiggsl\rightarrow 4\tau$, leads to a leptonic final state (with best fit around $\snmassr\approx25$~GeV), which differs from the usual $2\tau$ final state (whose best fit is around $10$~GeV~\cite{Calore:2014xka}).
We have also found $2\tau$ final states; however, these appear only for $\snmassr\lesssim5$~GeV \cite{Cerdeno:2014cda} and therefore fall out of the 95\% C.L. 
\\[1ex]

\noindent $\bullet$ $\snmassr\approx30 - 135$~GeV. 
This region is populated by points which present annihilation mainly into $b\bar{b}$ (grey), $c\bar{c}$ (green), $gg$ (violet), $\phiggsl\phiggsl$ (cyan), $\higgsl\higgsl$ (blue) and $\higgsl\higgsm$ (dark blue).

The best fit for a pure annihilation into a $b\bar{b}$ pair is obtained for $\snmassr=46.1$~GeV (see Table \,\ref{tab:chi}), in good agreement with Ref.~\cite{Calore:2014xka}, but it shifts to larger masses $\snmassr=63.1$~GeV if mixed final states are considered. 
Very few solutions with dominant $c\bar{c}$ and $gg$ final states are found. These channels dominate when the up component of the lightest Higgs is larger than the down component, which enhances the Higgs coupling to up-type fermions and top loop contributions to $gg$ final states. However, these loop contributions also enhance the $\gamma\gamma$ line production and most of the points are excluded for this reason. Besides, these final states are always related to the resonant annihilation of RH sneutrinos through a light singlet-like $\higgsl$~\cite{Cerdeno:2014cda} and typically have a smaller relic abundance  than the lower bound considered in this article. 
This also happens for other channels when $\snmassr\approx m_{H^0_2}/2\approx63$~GeV and explains the gap in the plot.

The annihilation into a pair of CP even Higgs bosons takes place mostly for $\snmassr\gsim60$~GeV. 
These subsequently decay mainly into $b\bar b$ (if the down component of $\higgsl$ is large), giving rise to a four quark final state $\snr_1\snr_1\rightarrow 2\higgsl\rightarrow 4 b$. It can also decay into gluons and $c\bar c$ (if the up component of $\higgsl$ is large). 
It is worth noting that this light $\higgsl$ is singlet-like, typically has a mass of $30-100$~GeV, 
very typical of the NMSSM. For this reason the best fit point (which in our analysis is obtained for $\snmassr\approx121$~GeV with $\chi^2=21.6$) differs from other solutions found for the SM Higgs \cite{Calore:2014nla}.

Finally, regarding the annihilation into $\phiggsl\phiggsl$, the pseudoscalar in this region satisfies $\phmassl>2m_b$ and thus decays predominantly into $b\bar{b}$. The annihilation into four $b$ quarks, $\tilde{N}_1\tilde{N}_1\rightarrow 2\phiggsl\rightarrow 4 b$, produces a best fit of the GCE for $\snmassr=65.0$~GeV when this channel is almost pure. 
Mixed scenarios are in fact responsible for the best fit point found in our analysis with 44\% of annihilation into $\phiggsl\phiggsl$, a RH sneutrino mass $\snmassr=63.5$~GeV  and a $\chi^2=20.8$.\\[1ex]

The spectra for the best fit points in Table \,\ref{tab:chi} are shown in Figure~\ref{fig:spec}. As we can see the pseudoscalar and scalar channels provide the best fits.
This is mostly due to the presence of spectral features (mainly lines in the case of $\phiggsl\phiggsl$ and box-shaped emissions for $\higgsl\higgsl$) that improves the fit at large energies. 
On top of this, the energy distribution of the spectra of four-body final states (fermions, gauge bosons or combinations between these two, as shown in Fig.1) is also known to provide better fits to the GCE \cite{Martin:2014sxa}.
Other two-body final states, such as $b\bar{b}$,  $c\bar{c}$,  $gg$ and  $W^+W^-$, also benefit from the presence of lines in the spectrum
which might be probed by Fermi-LAT using the latest Pass 8 reprocessed data.

In Figure~\ref{fig:sv95P7}, we also show all the points that are not ruled out by our estimation of the Pass~8 dSph bounds and provide a 95\% C.L. fit to the GCE.
It is remarkable that all the solutions found that fit the GCE at 68\% C.L. would be now ruled out, highlighting the tension between the Pass 8  data and the GCE.
Moreover, the points that survive the Pass~8 constraints are mainly related to dominant annihilation into non standard final states, such as $A^0_1A^0_1$ and $H^0_1H^0_1$. 
This is due to the fact that these solutions produce four particles in the final state, which makes their spectrum different from that of the usual two-body final states.   
Furthermore, at the same time they produce spectral features at high energies. 
It is worth noting that the points which are still alive and fit the GCE at 95\% C.L. might show up in the results of the Fermi-LAT Collaboration in the very near future.

\begin{figure*}[t!]
      \begin{center}
\scalebox{0.9}{
            \includegraphics[scale=1]{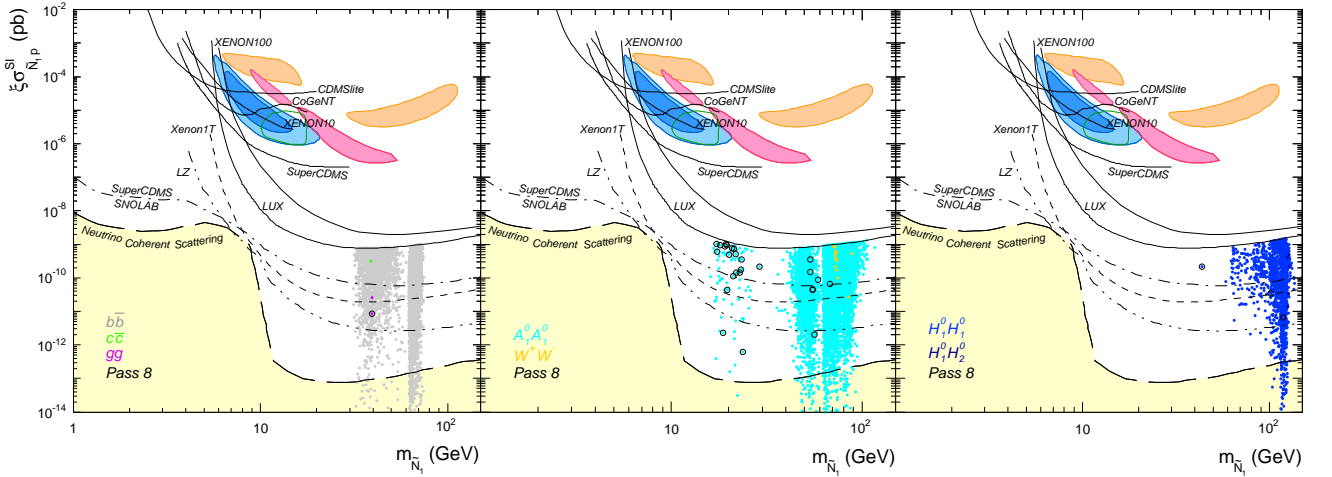}
               }
	\caption{\footnotesize Theoretical predictions for $\crosssec$ as a function of $\snmassr$ for points which fit the GCE at 95\% C.L. and fulfil all the experimental bounds. The colour convention is as in Fig.\,\ref{fig:sv95P7}. Solid lines represent the current  experimental upper bounds from direct detection experiments, whereas dotted lines are the projected sensitivities of next-generation detectors. The dashed line corresponds to an approximate band where neutrino coherent scattering with nuclei will begin to limit the sensitivity of direct detection experiments. Closed contours represent the areas compatible with the observed excesses in  DAMA/LIBRA (orange), CRESST (red), CDMS II (blue), and CoGeNT (green).}
      \label{fig:si95P7}
      \end{center}
\end{figure*}

\subsection{Implications for direct DM searches}

The RH sneutrino elastic scattering off quarks is mediated by the exchange of a CP-even Higgs boson on a $t$ channel, and the dominant contribution arises from the lightest of these particles, $H^0_1$.
As it was shown in Ref.~\cite{Cerdeno:2011qv}, 
the annihilation cross section and the spin-independent scattering cross section off protons, $\crosssec$, are correlated by crossing symmetry
when the main annihilation final state in the Early Universe is any quark pair.
This generically leads to a large $\crosssec$, already excluded by current direct detection limits. 
However, this correlation can be broken in the presence of resonant annihilations.
Because of this as well as the different annihilation final states, the predictions for $\crosssec$ span many orders of magnitude~\cite{Cerdeno:2014cda}.

As already emphasised in the previous section, there is a wide variety of final states that can fit the GCE at 95\% C.L.; therefore the prospects for direct detection can be very different.
In Figure~\ref{fig:si95P7}, we show the theoretical predictions for $\xi\sigsi$ as a function of the RH sneutrino mass for all the points that fit the GCE at 95\% C.L. 
We have also included the most stringent bounds from direct detection experiments \cite{Akerib:2013tjd,Agnese:2013jaa,Agnese:2014aze} and some of the predicted sensitivities of next-generation detectors.

It is noteworthy that many of the points have a relatively large scattering cross section and might be within the reach of future experiments, such SuperCDMS and LZ. In particular, this is the case of most of the examples with $A^0_1A^0_1$ final states and $\snmassr\gsim30$~GeV that also satisfy our estimation of the Pass 8 constraints. 
On the contrary, especially in the points with resonances, the predicted $\xi\crosssec$ can be extremely small. 
The resonance with the SM-like Higgs $\higgsm$ is narrow and occurs around $\snmassr\approx\hmassm/2\approx 62.5$~GeV. The resonance with the lightest singlet-like Higgs, $\higgsl$, seems broader due to the variation of $\hmassl$ throughout the scan and generically occurs for $\snmassr\lesssim60$~GeV. 
Fortunately, the presence of such light scalar and/or pseudoscalar Higgs bosons in these cases might give rise to interesting signals at the LHC.

\subsection{Implications for indirect DM searches}

\begin{figure}[t!]
      \begin{center}
\scalebox{0.9}{
            \includegraphics[scale=0.4]{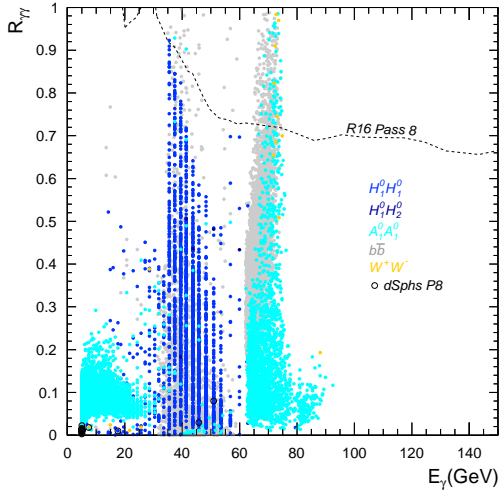}
               }
\caption{\footnotesize $R_{\gamma\gamma}$ ratio as a function of the photon energy, $E_\gamma$,  for which the ratio is maximised.
All the points fulfil all the experimental bounds. The colour convention is as in Fig.\,\ref{fig:sv95P7}.
The dashed line denotes our estimation of the improved sensitivity to spectral feature searches with Fermi-LAT Pass~8 data.
}
      \label{fig:R}
      \end{center}
\end{figure}

As already pointed out, most of the cases that provide good fits to the GCE also contain box-shaped features and lines in the gamma-ray spectrum. 
Although we have included in our analysis the constraints from the Fermi-LAT Collaboration on these contributions, 
future searches for these kind of features might provide new insight on the DM properties. In this subsection, we will determine which of the points that fit the GCE at 95\% C.L. are also within the reach of future searches for gamma-ray lines.

To this aim, we have defined the following ratio:
\begin{equation}
	R_{\gamma\gamma}=\frac{\xi^2\langle\sigma v\rangle_{sf}(E_\gamma)}{\langle\sigma v\rangle_{\gamma\gamma}^{LAT}(E_\gamma)/r}\ ,
\end{equation}
which quantifies how far our predictions are from the current experimental limits.
Here $\xi^2\langle\sigma v\rangle_{sf}$ is the contribution from our model to spectral features (box or lines) on the gamma-ray spectrum and $\langle\sigma v\rangle_{\gamma\gamma}^{LAT}$ is the expected limit from Ref.~\cite{Ackermann:2013uma} for R41. 
The factor $r\approx1.52$ has been introduced in the previous section to convert the bounds of Ref.~\cite{Ackermann:2013uma} to the DM halo considered in this article.
$R_{\gamma\gamma}$ is evaluated at the energy $E_\gamma$, for which the ratio is maximised.
For gamma-ray lines $E_\gamma$ coincides with the energy of the line. On the other hand, for box-shaped features 
$E_\gamma$ represents the mean value of the Fermi-LAT energy bin for which $\xi^2\langle\sigma v\rangle_{sf}$ is closer to the current expected limit.
Notice that $R_{\gamma\gamma}<1$, since the current bound on spectral features has already been applied to our data.

In Figure~\ref{fig:R}, we represent $R_{\gamma\gamma}$ vs. $E_\gamma$ for all the points that fit the GCE at 95\% C.L.  
We also indicate with a dashed line the expected improvement on these kind of searches with the Pass~8 data\footnote{This estimation is obtained assuming that the improvement on the expected limit 
with respect to the Pass~7 data for the Einasto profile (ROI R16) shown in Ref.~\cite{fermi-symp-lines} can also be applied to the DM halo considered in this article.}.
We can observe that many of these scenarios have $R_{\gamma\gamma}>0.5$, which means that an improvement of a factor 2 in the Fermi-LAT sensitivity to the search for 
spectral features would be enough to probe these solutions.

As has already been emphasised, the points in which the RH sneutrino annihilates mainly into scalar or pseudoscalar Higgs bosons typically 
present box-shaped features and/or lines in their spectrum. The energy $E_\gamma$ at which we evaluate $R_{\gamma\gamma}$ (and at which we expect them to be detected) is systematically shifted towards low values, 
since the sensitivity of the Fermi-LAT instrument is better\footnote{When both lines and box-shaped features are present in the same spectrum, the better sensitivity of the Fermi-LAT to low-energies implies that the optimal value of $E_\gamma$ is typically achieved close to the lower-end of the box.}.
In particular, all the points with $\higgsl\higgsl$ final states have $E_\gamma<60$ GeV (in spite of the RH sneutrino mass being $\snmassr\approx60-135$~GeV). 
Regarding the points with $\phiggsl\phiggsl$ final states, only a portion of them with $E_\gamma\approx 70$~GeV might be observable in the near future.

Unfortunately, all of the points that satisfy our expectation of the Pass 8 bound on dSphs have very small values of $R_{\gamma\gamma}$ and therefore would not present any observable feature in the gamma-ray spectrum.

\subsection{Implications for LHC searches}

\begin{figure}
      \begin{center}
\scalebox{0.9}{
             \includegraphics[scale=0.4]{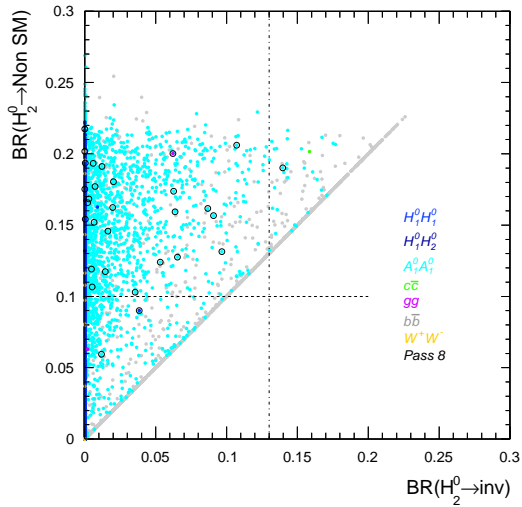}

               }
\caption{\footnotesize Branching ratio of the SM-Higgs, $H^0_2$, into non SM particles as a function of its invisible branching ratio. 
All the points fulfil all the experimental bounds.
The colour convention is as in Fig.\,\ref{fig:sv95P7}.
The regions above the dashed line and to the right of the dot-dashed line will be probed by future searches at the LHC.}
      \label{fig:higgs}
      \end{center}
\end{figure}

Finally, in this section we investigate the implications of new physics searches at the LHC. In general, the NMSSM introduces profound changes in the structure of the Higgs sector. Most notably, the addition of new states opens the door to the occurrence of light singlet-like scalar and pseudoscalar Higgs bosons with a mass that can be considerably smaller than that of the SM Higgs.
As we have seen in the previous sections, these scenarios are very common when we try to fit the GCE with RH sneutrino DM at 95\% C.L. 
These light particles provide new decay channels for the SM Higgs boson, $\higgsm$. 
Some of them contribute to the invisible branching fraction, as is the case of the RH sneutrino, the RH neutrino and the lightest neutralino ($\higgsm\to\snr_1\snr_1$, $\higgsm\to\rhn\rhn$ and $\higgsm\to\neutl\neutl$), 
whereas others would only appear as non-standard decay channels\footnote{Recall that the invisible decay modes are also included in the non-standard branching ratio.}($\higgsm\to\higgsl\higgsl$ and $\higgsm\to\phiggsl\phiggsl$).

Current LHC data leave considerable room for these contributions [e.g., BR$(\higgsm\to {\rm inv})<0.27$]. Nevertheless, the future high luminosity LHC run (3000 fb$^{-1}$ at $\sqrt{s}=14$~TeV) will significantly narrow down these ranges 
(e.g., the invisible branching ratio in $ZH$ associated production will be probed at 95\%C.L. down to 13\% and the Higgs couplings to SM particles will be measured individually with 10\% accuracy~\cite{LHC-Higgs}). 
It is therefore conceivable that some of the scenarios considered in this article can be probed in this way.

We have defined the invisible branching ratio of the SM Higgs and the branching ratio of the SM Higgs into non SM particles as follows,
\begin{eqnarray}
	{\rm BR}(\higgsm\to {\rm inv}) &=& {\rm BR}(\higgsm\to\snr_1\snr_1)\nonumber\\
	&&+{\rm BR}(\higgsm\to\rhn\rhn) \nonumber\\
	&&+{\rm BR}(\higgsm\to\neutl\neutl)\nonumber\\ 
	{\rm BR}(\higgsm\to {\rm Non\ SM}) &=& {\rm BR}(\higgsm\to\higgsl\higgsl)\nonumber\\
	&&+{\rm BR}(\higgsm\to\phiggsl\phiggsl)\nonumber\\ 
	&&+{\rm BR}(\higgsm\to {\rm inv})\, .
\end{eqnarray}
Mostly only the RH sneutrinos contribute to the invisible branching fraction, since the RH neutrino mass is generally too large.
Neutralinos in this scenario would decay into the RH sneutrino through $\neutl\to\snr_1\rhn$ or $\neutl\to\snr_1\nu_L$ \cite{Cerdeno:2013oya}, contributing to the invisible branching ratio of the SM Higgs boson. Nevertheless, this contribution is negligible in most of the points of the parameter space since either the lightest neutralino is too heavy or the branching ratio of the $\higgsm\to\neutl\neutl$ process is very small.

In Figure~\ref{fig:higgs}, we plot ${\rm BR}(\higgsm\to {\rm Non\ SM})$ as a function of ${\rm BR}(\higgsm\to {\rm inv})$ 
for all the points that fit the GCE at 95\% C.L. 
We denote with dashed and dot-dashed lines the expected sensitivity in the future high luminosity LHC run to these two observables, respectively.
Many solutions found here would lead to observable signals, mainly in searches of non SM decays of $\higgsm$. 
Interestingly, this is the case of most of the solutions to the GCE that survive our estimation of the Pass 8 constraints on dSphs, which have  ${\rm BR}(\higgsm\to {\rm Non\ SM})>0.1$. 
Furthermore, the $A^0_1A^0_1$ final states that fulfil the Pass 8 bounds from dSphs might be also probed at the LHC, through the direct production of $A^0_1$ and its subsequent decay into $\tau$ leptons producing a multilepton signal~\cite{Cerdeno:2013cz,Cerdeno:2013qta}.

\begin{figure*}
      \begin{center}
\scalebox{0.9}{
            \includegraphics[scale=0.4]{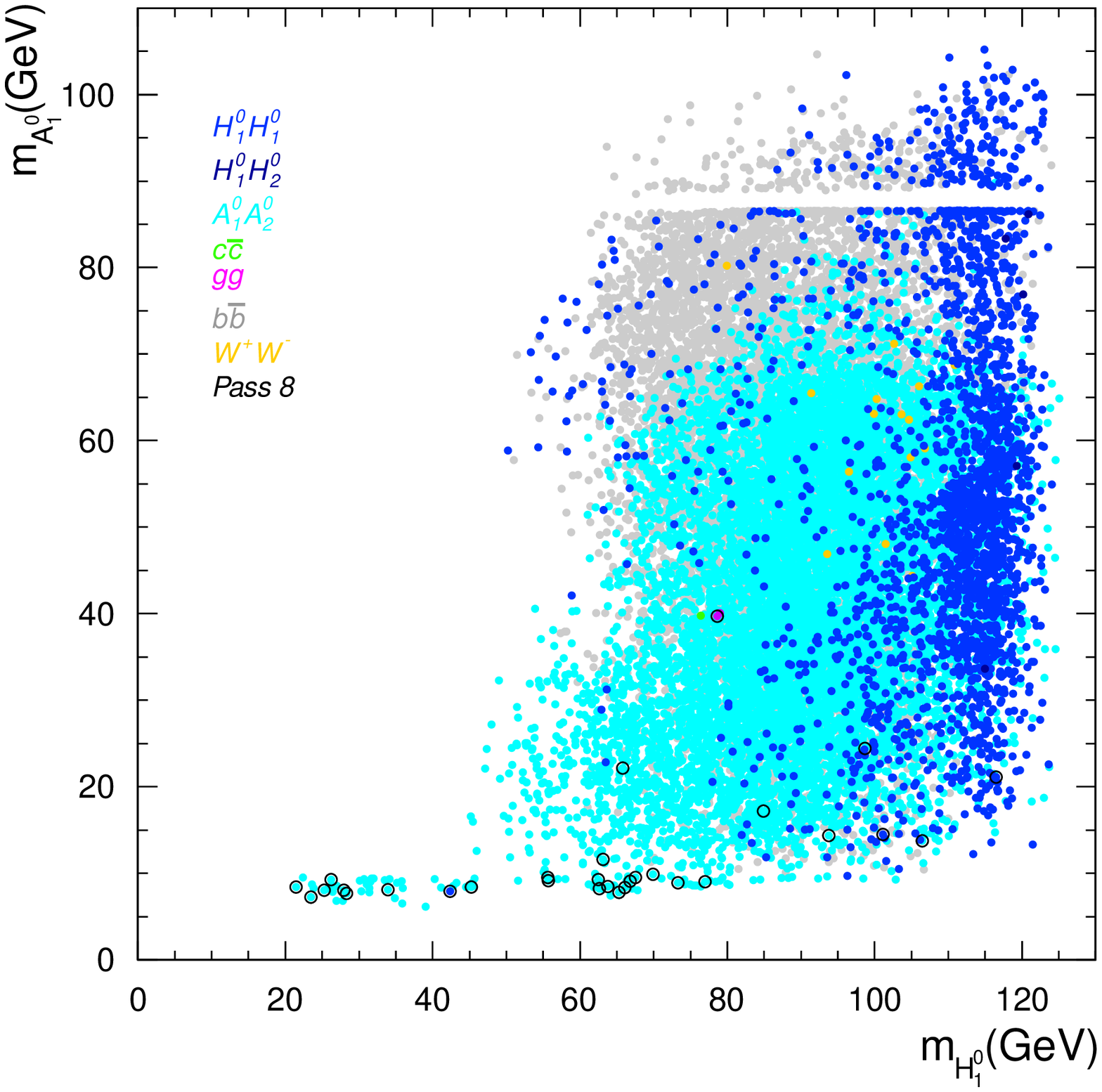}
            \includegraphics[scale=0.44]{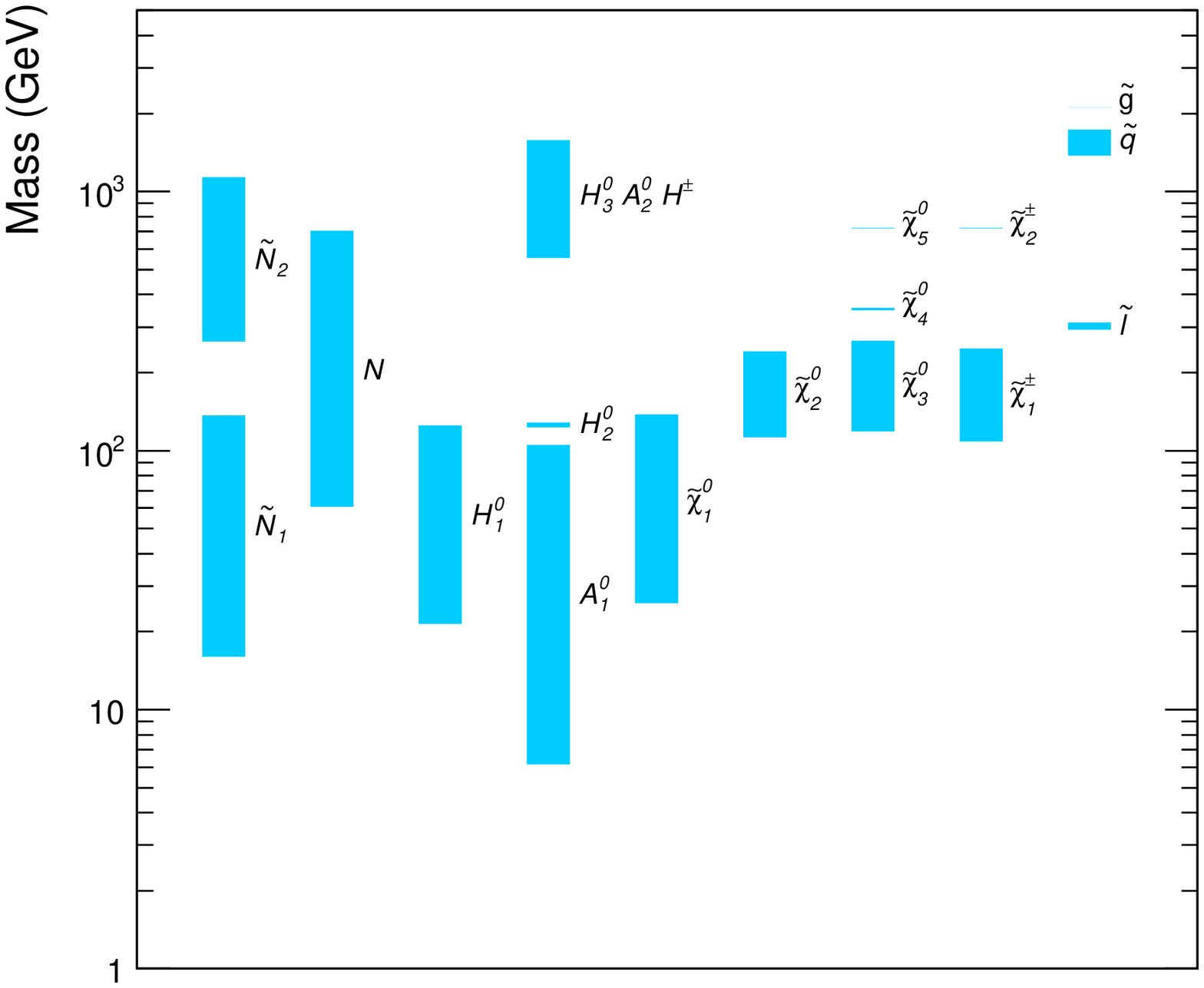}
                           }
	\caption{\footnotesize Left: $m_{A^0_1}$ as a function of $m_{H^0_1}$ for the points fulfilling all the experimental constraints and fitting the GCE at 95\% C.L. with the colour convention as in Fig.\,\ref{fig:sv95P7}. Right: Mass ranges for the Higgs sector and the different supersymmetric particles corresponding to the viable points in the parameter space.}
      \label{fig:masses}
      \end{center}
\end{figure*}

\section{Discussion of the viable parameter space}

As seen previously, some of the best fits to the GCE in this model require the presence of either a light CP-even or a CP-odd Higgs boson (see Table~\ref{tab:chi}). In fact, after the inclusion of the dSph Pass 8 bounds, only the regions with a very light singlet-like CP-odd Higgs boson survive.

In Figure~\ref{fig:masses} (left panel), we represent the solutions that fulfil all the experimental constraints and fit the GCE at 95\% C.L. in the plane $(m_{A^0_1},\, m_{H^0_1})$. In general, points with light CP-even Higgs bosons below the mass threshold $m_{H^0_2}/2$ are more difficult to obtain than those with a light CP-odd Higgs\footnote{
This is a consequence of the coupling of these Higgses to $H^0_2$ (the SM-like Higgs). Our choice of signs for $\lambda$ and $\kappa$ makes the coupling of $H^0_1$ to $H^0_2$ higher than the $A^0_1$ one; 
thereby the existing LHC constraints on the SM-like Higgs couplings to SM particles are more stringent for the $H^0_1$ final states~\cite{Cerdeno:2013qta}. This does not apply to the limit $\kappa\to 0$, where the relative sign between $\lambda$ and $\kappa$ does not play any role.}.  
The mass of the pseudoscalar Higgs can even be smaller than $2m_b$ (thus favouring its decay into leptons), whereas the mass of the lightest scalar Higgs boson remains above 20~GeV. It is noteworthy that the solutions allowed by the estimated Pass 8 bounds from dSphs always contain a pseudoscalar Higgs below 50~GeV, and some of them also include a scalar Higgs below 60~GeV. 

In Figure~\ref{fig:masses} (right panel), we show the mass ranges of the different supersymmetric particles, as well as the Higgs sector for the aforementioned solutions. In general, these solutions have a mass spectrum that contains several non SM particles below the frontier of 100 GeV. Namely, we can see that neutralinos in these scenarios can be as light as 20~GeV while in some cases also the RH neutrinos can reach a mass around 60~GeV. Notice that the masses of sleptons and squarks do not vary much, since the soft masses and trilinear parameters for these particles were fixed in our scan.

\begin{figure*}
      \begin{center}
\scalebox{0.9}{
            \includegraphics[scale=0.4]{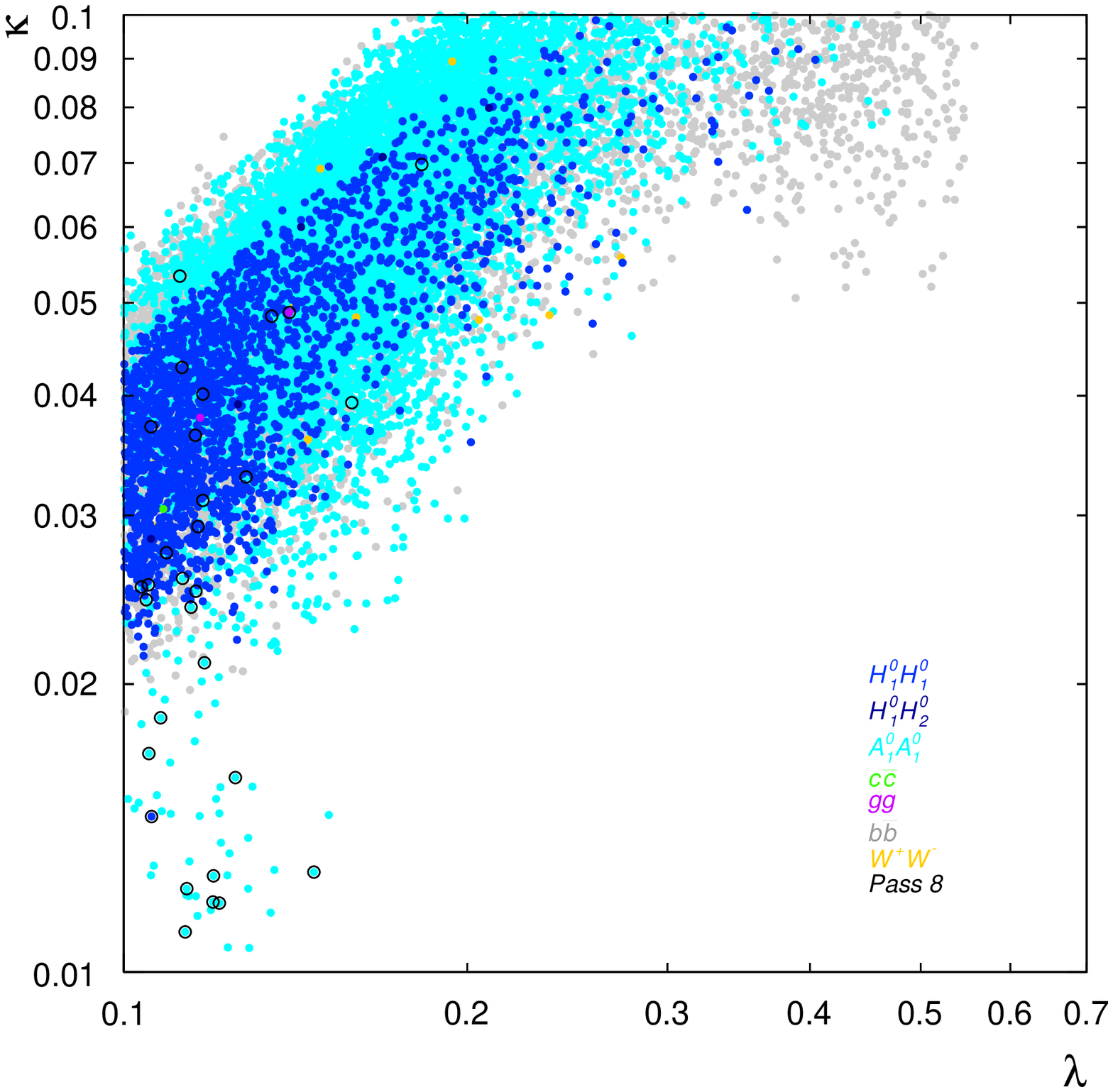}
            \includegraphics[scale=0.4]{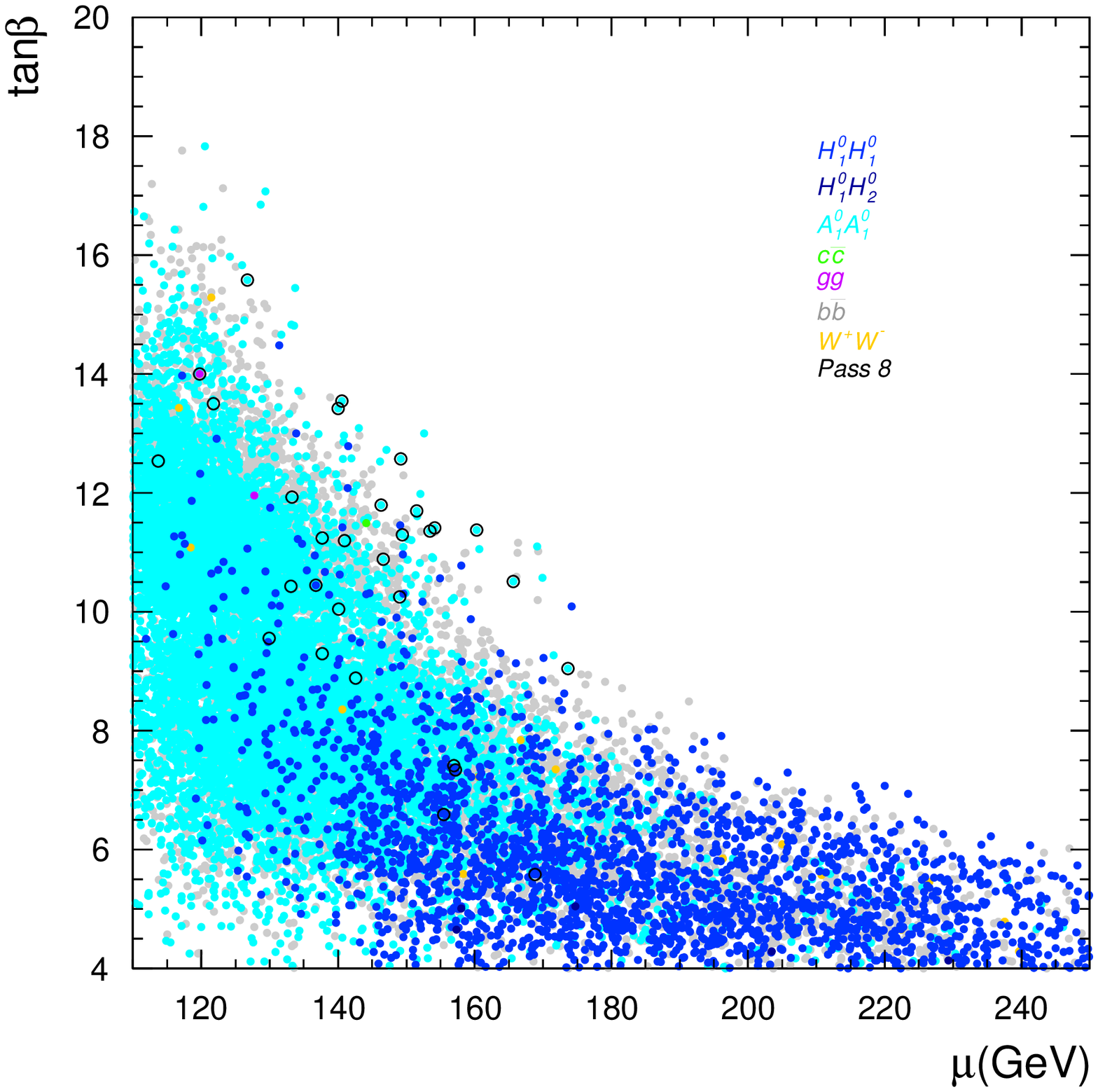}
               }
	\caption{\footnotesize Left: $\kappa$ versus $\lambda$ for the points that fulfil all the experimental constraints and fit the GCE at 95\% C.L. with the colour convention as in Fig.\,\ref{fig:sv95P7}. Right: The same but for the $\tan\beta$ and $\mu$ parameters.}
      \label{fig:inputs}
      \end{center}
\end{figure*}

Figure~\ref{fig:inputs} displays the solutions fitting the GCE at 95\% C.L. in the planes $\kappa$ versus $\lambda$ (left panel) and $\tan\beta$ versus $\mu$ (right panel), where the main constraint on these parameters is the bound on the $H^0_2$ mass~\cite{Cerdeno:2014cda}.
For $H^0_1H^0_1$ final states (blue) we observe a slight preference for small values of $\lambda$, while for $A^0_1A^0_1$ (cyan) and $b\bar{b}$ (grey) final states the solutions can be found for a wide range of values. 
Very light pseudoscalar Higgses can be found in the Peccei-Quinn (PQ) limit $\kappa\rightarrow0$ and $A_\kappa\rightarrow0$, for which the $U(1)_{PQ}$ symmetry is restored, and it is precisely in these regions where we find most of the points that are in agreement with the expected dSph bounds using Pass~8 data.
In the right panel of Fig.~\ref{fig:inputs}, we can observe that these solutions are grouped in the region of $\mu<180$~GeV  while the values of $\tan\beta$ do not show any preferred region.
The PQ limit is not needed for points that only fulfil Pass~7 data as the pseudoscalar mass can be larger.

\subsection{Comparison with low-mass neutralino DM in the NMSSM}

\begin{figure}
      \begin{center}
\scalebox{0.9}{
            \includegraphics[scale=0.4]{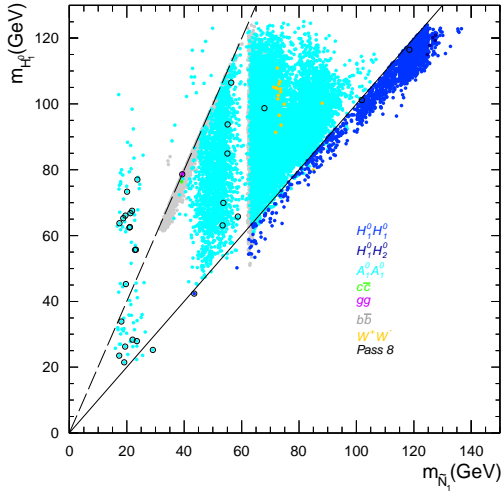}
               }
	\caption{\footnotesize $m_{H^0_1}$ versus $m_{\tilde{N}_1}$ for the points fulfilling all the experimental constraints and fitting the GCE at 95\% C.L. with the colour convention as in Fig.\,\ref{fig:sv95P7}. The solid line corresponds to $m_{H^0_1}=m_{\tilde{N}_1}$ while the dashed line denotes the resonant annihilation condition, $m_{H^0_1}=m_{\tilde{N}_1}/2$.}
      \label{fig:mh1-msn}
      \end{center}
\end{figure}

The NMSSM can also accommodate low-mass neutralino DM~\cite{Gunion:2005rw,Ferrer:2006hy,Aalseth:2008rx,Vasquez:2010ru,Cao:2011re,AlbornozVasquez:2011js,Kozaczuk:2013spa}. Several studies have recently pointed out that these neutralinos can also explain the GCE~\cite{Cheung:2014lqa,Huang:2014cla,Gherghetta:2015ysa} when they are either singlino/Higgsino or bino/Higgsino admixtures~\cite{Cheung:2014lqa}. 
The former situation can occur when $\kappa/\lambda\ll1$ and requires some degree of fine-tuning in the Higgs mass matrix parameters in order to avoid large deviations from the SM-like Higgs composition. Moreover, to account for the observed relic abundance the mass of the pseudoscalar Higgs must be very close to the resonant condition, $\phmassi\approx 2\,m_{\tilde{\chi}^0_1}$. The bino/Higgsino 
scenario requires $\kappa/\lambda\gg1$ and is however less fine-tuned since no resonant condition is required to fulfil the relic density constraint. 

The region of the parameter space in which RH sneutrinos can explain the GCE is similar to that with singlino/Higgsino neutralinos, as $\kappa/\lambda<1$ (in fact, light neutralinos are also frequent as we showed in Fig.\,\ref{fig:masses}). 
It is worth noting that, especially when Pass~8 constraints are imposed, both models feature a similar Higgs spectrum with light singlet-like scalar and pseudoscalar Higgs bosons. 
Nevertheless, in the NMSSM with RH sneutrinos the amount of fine-tuning required is largely reduced as resonant annihilation is no longer necessary in order to obtain the correct relic abundance.
In this model, the annihilation cross section increases when the annihilation channels into light scalar and pseudoscalar Higgs bosons ($\snr_1\snr_1\to\higgsl\higgsl$ and $\snr_1\snr_1\to\phiggsl\phiggsl$) are open \cite{Cerdeno:2014cda}. This is illustrated in Figure~\ref{fig:mh1-msn}, where $m_{H^0_1}$ is represented as a function of $m_{\tilde{N}_1}$ for the set of points for which the GCE is fit at 95\% C.L. The solid line denotes $m_{H^0_1}=m_{\tilde{N}_1}$ (threshold for the annihilation in $H^0_1H^0_1$) and the dashed line corresponds to $m_{H^0_1}=m_{\tilde{N}_1}/2$ (condition for resonant annihilation\footnote{We remind the reader that the RH sneutrino does not couple to the light pseudoscalar.}). Only the points with $b\bar b$ final states require resonant annihilation.

It should finally be emphasized that, contrary to the neutralino case, the parameters that determine the RH sneutrino properties (the soft mass, $\snmassr$, the trilinear term, $\aln$, and the coupling, $\ln$) do not affect the masses of the Higgs sector of the NMSSM. Therefore, it is much easier to obtain the correct relic abundance without violating the LHC bounds.

\section{Conclusions}

In this article, we have demonstrated that RH sneutrino dark matter in the NMSSM can account for the observed low-energy excess in the Fermi-LAT gamma-ray spectrum from the Galactic Centre. 
Since we are working with a complete theoretical framework, we have also explored the complementarity with other DM search strategies, such as indirect searches for gamma-ray lines, direct DM detection, and the implications for a future LHC run.

More specifically, we have performed a scan in the parameter space of the model, incorporating all the constraints from the LHC as well as direct and indirect dark matter searches. For the latter, we have included the Fermi-LAT bounds on dSphs and gamma-ray spectral features. In addition, we have estimated the effect of the latest Fermi-LAT reprocessed data (Pass 8).
We have computed the gamma-ray spectrum for each point in the parameter space of the model, and then we have performed a $\chi^2$ fit to the excess.
It should be emphasised that, contrary to usual model-independent approaches, we have taken into account the full annihilation products, without assuming pure annihilation channels.

We have obtained good fits to the GCE for a wide range of the RH sneutrino mass, $\snmassr\approx20-135$~GeV, with annihilation cross section in the DM halo $\xi^2\sigmav\approx5\times10^{-27}-8\times10^{-26}$cm$^3$/s at 95\% C.L. There is a large variety of final states for the RH sneutrino annihilation. In general, we observe that the fit to the GCE is good when the RH sneutrino annihilates mainly into pairs of light singlet-like scalar or pseudoscalar Higgs bosons, $\higgsl\higgsl$ or $\phiggsl\phiggsl$, that subsequently decay in flight. These produce spectral features, such as gamma-ray lines and/or box-shaped emissions that improve the goodness of the fit at large energies. 
The best fit of our analysis ($\chi^2=20.8$) corresponds to a RH sneutrino with a mass of $64$~GeV which annihilates preferentially into a pair of light singlet-like pseudoscalar Higgs bosons (with masses of order 60 GeV). Good fits are also obtained for a variety of mixed final states and have been summarised in Table\,\ref{tab:chi}.

Our estimation of the constraints from the new Pass~8 data from Fermi-LAT rules out all the points that fit the GCE at 68\% C.L. and leaves only a few of them within the 95\% C.L. region. Interestingly, these correspond mainly to points with predominant annihilation into light  pseudoscalar Higgs bosons. It is remarkable that a region with $\snmassr\approx25$~GeV remains valid.In this region, the light pseudoscalar has a small mass, $\phmassl<2m_b$, and therefore leads to a leptonic final state with 4$\tau$, $\snr_1\snr_1\rightarrow 2\phiggsl\rightarrow 4\tau$, which differs from the usual result for $2\tau$.

We have then studied the implications for direct detection of the points that fit the GCE at 95\% C.L. Because of the presence of light non SM particles in the mass spectrum of the model, and the fact that many of our solutions involve resonant annihilation through the lightest scalar Higgs, the prospects for the RH sneutrino scattering cross section off quarks span several orders of magnitude.
It is remarkable that many of the points found here are within the reach of next-generation experiments such as SuperCDMS and LZ. Interestingly, this is the case of the few points which also survive our estimation of the Pass~8  constraints.

We have also investigated the prospects for the detection of these scenarios in future searches for gamma-ray spectral features, using the estimated future sensitivity of Fermi-LAT for this kind of studies. We have found that only a small fraction of the points that fit the GCE at 95\% C.L. will be observable in this way, mainly those associated with resonant annihilation through the lightest or second-lightest scalar Higgs boson. Unfortunately, none of the scenarios allowed by our estimation of the Pass 8 constraints would be detectable by spectral feature searches.
  
Finally, we have computed the contribution of these scenarios to the invisible and non-SM branching ratios of the SM Higgs and compared them with the predicted reach of the future high-luminosity LHC run at 14 TeV. We have determined that a substantial amount of viable scenarios (many of which also satisfy the Pass~8 constraint) can be probed in such a way. 

It is remarkable that many of the scenarios that can explain the GCE with RH sneutrino DM can be subject to a complementary detection through other methods in the near future, such as direct DM searches or collider tests of the Higgs sector.

\subsection{Acknowledgements}
The authors are grateful to V. Gammaldi, G. A. Gomez-Vargas, D. Hooper, T. Linden, R. Lineros, and C. Weniger  for useful comments. 
D.G.C. is partially supported by the STFC, and S.R. by the Campus of Excellence UAM+CSIC.
We thank the support of the Consolider-Ingenio 2010 program under Grant No. MULTIDARK CSD2009-00064, the Spanish MICINN under Grants No. FPA2012-34694 and FPA2013-44773-P, the Spanish MINECO Centro de Excelencia Severo Ochoa Program under Grant No. SEV-2012-0249,
and the European Union under the ERC Advanced Grant SPLE under Contract ERC-2012-ADG-20120216-320421.

\end{document}